\documentclass[preprint,showpacs,amsmath,amssymb,endfloats]{revtex4}

\usepackage{graphicx}
\usepackage{dcolumn}
\usepackage{bm}
\usepackage{epstopdf}
\usepackage{latexsym}
  
\newcommand{\be}{\begin{equation}}
\newcommand{\ee}{\end{equation}}
\newcommand{\bea}{\begin{eqnarray}}
\newcommand{\eea}{\end{eqnarray}}

\newcommand{\req}[1]{Eq.~(\ref{#1})}

\begin{document}


\title{Spin-Orbit Scattering and Quantum Metallicity in Ultra-Thin Be Films} 

\author{Y.M.  Xiong, A.B. Karki, D.P. Young, and  P. W. Adams}
\affiliation{Department of Physics and Astronomy\\Louisiana State University\\Baton Rouge, Louisiana, 70803}%


\date{\today}

\begin{abstract}
 We compare and contrast the low temperature magnetotransport properties of ultra-thin, insulating, Be films with and without spin-orbit scattering (SOS).  Beryllium films have very little intrinsic SOS, but by ``dusting'' them with sub-monolayer coverages of Au, one can introduce a well controlled SOS rate.  Pure Be films with sheet resistance $R>R_Q$ exhibit a low-temperature negative magnetoresistance (MR) that saturates to the quantum resistance $R_Q=h/e^2$.  This high-field {\it quantum metal} phase is believed to represent a new ground state of the system.  In contrast, the corresponding negative MR in Be/Au films is greatly diminished,  suggesting that, in the presence of strong SOS, the quantum metal phase can only be reached at field scales well beyond those typically available in a low temperature laboratory. 
\end{abstract}

\maketitle
 
 Ultra-thin metal films have proven to be extraordinarily fertile systems for studying a variety of quantum scattering and interaction processes that ultimately serve to destroy the metallic state of their bulk counterparts \cite{BergmannWL,Dynes1978}.  By the early 1980's it was recognized that coherent backscattering in moderately disordered films produces logarithmically insulating behavior at low temperature \cite{LeeRam}.  In addition, disorder tends to enhance the impact of electron-electron ($e-e$) interactions, which manifest themselves as a logarithmic suppression of the density of states near the Fermi energy \cite{AltshulerRev}.  The theoretical description of weakly disordered two-dimensional systems has, in fact, been a great success, having given us a quantitative description of a wide spectrum of transport and tunneling density-of-states experiments \cite{gang4,LeeRam}.  In contrast, the magnetotransport properties of highly disordered films, with sheet resistance R greater than the quantum resistance $R_Q=h/e^2$, remain poorly understood \cite{Butko2001}.  To date there is no clear consensus as to what roles film morphology \cite{Epstein1983}, phase coherent hopping \cite{Nguyen1985,Wohlman1989,Medina1996}, Zeeman splitting \cite{Goldman1,Matveev1995,Butko2000}, and/or spin-orbit scattering \cite{Shapir1989,Hernandez1992,Pichard1990} play in producing the correlated insulator phase of ultra-thin metal films.   Recently, however, investigators have recognized that new insights into the processes that contribute to the formation of the correlated insulator phase can be obtained through the study of metal films that undergo a superconductor-to-insulator (S-I) transition \cite{Goldman2}.  The reason for this is obvious.  On the one hand, superconductors are characterized by a macroscopic quantum state which exhibits long range phase coherence and non-dissipative current flow.  Insulators, on the other hand, have no long range coherence of any sort, and exhibit dissipative, glassy dynamics.  The fact that this striking juxtaposition of electronic properties can be controlled via an external tuning parameter, such as film resistance, allows one to explore the emergence of the insulating phase from the perspective of the superconducting phase and its attendant fluctuations.   
 
 In practice, a superconducting film can be driven into the insulating phase by increasing its disorder beyond a specific threshold.  Typically this is done by making the film thinner, and, once the normal state sheet resistance is of the order of $R_Q$, the superconducting phase gives way to a highly insulating phase \cite{Goldman3}.  Alternatively, if one is close to the insulating threshold, a magnetic field can be used to tune the system through the S-I transition \cite{Hebard}.  The S-I transition has been the subject of intense investigation for more than two decades now, but the primary interest has been in those systems which are homogeneously disordered, and, in particular, non-granular.  It is generally believed that the disorder-driven SI transition in these systems is mediated by $e-e$ interaction effects \cite{Valles1}.  With increasing disorder, an otherwise perturbative depletion of quasiparticle states at the Fermi energy, grows into a full blown correlation gap when $R>>R_Q$ \cite{Butko2000b,Butko2001}.  This has the effect of undermining the superconducting order parameter amplitude, thereby suppressing the transition temperature $T_c$ \cite{Valles1}.  The exact nature of the insulating state, and its relation to the superconducting phase remains unclear.  For instance, anomalously large, multi-fold, negative magnetoresistances have been reported in ultra-thin TiN films \cite{Baturina2007},  InO$_x$ films \cite{Shashkin1998,Gantmakher2000,Kapitulnik2005} and in insulating Be \cite{Butko2000b} films.  The MR of these films saturates at a weakly temperature dependent resistance that is always near $R_Q$, {\it i.e.} the ``quantum metal'' phase \cite{Baturina2007,Butko2001}. This observation suggests that the zero-field insulating ground state is distinctly different from the high-field ground state.  This has led to speculation that the zero-field ground state has an incoherent superconducting component \cite{Shahar,Valles2}.   In this Letter, we investigate the effect of spin-orbit scattering (SOS) on the MR behavior of Be films, and, their corresponding high-field quantum metal phase.  Interestingly, both Be and TiN films form dense homogeneously disordered, non-granular, films with an intrinsic, clean limit, $T_c\sim1$ K.   As we show below, the fact that both of these systems have a well documented, low SOS rate is crucial to the observation of the quantum metal phase.
 
 Numerous studies of the spin-paramagnetic transition in ultra-thin Be and Al films have shown that these light elements have a very low intrinsic SOS rate \cite{Tedrow1979,Adams2004,Adams1998} and are true spin-singlet superconductors.   However, it was initially shown by Tedrow and Meservey \cite{Tedrow1982} that a controllable amount of SOS could be induced in thin Al films by coating them with heavy noble metals.   In particular, they reported that, for each monolayer of Pt deposited on a 40 \AA-thick Al film, the SOS scattering rate $h/\tau_{so}$ increased by 3.2 meV.  Similarly,  large SOS rates can be induced in Be films by coating them with Au \cite{Wu2006}.   The most pronounced effect of SOS is to disrupt the spin rotation symmetry of the system, so that spin is no longer a good quantum number.  However, conventional BCS superconductivity does not require spin rotation symmetry, therefore SOS has little effect on the zero field properties of the condensate.  Nevertheless, the spin response of a superconductor, as probed by a parallel magnetic field, is much different in the presence of  SOS.   Indeed, Ref.\ \onlinecite{Wu2006} reported spin-paramagnetically limited critical fields that were almost an order of magnitude higher than Clogston-Chandrasekhar limit in thin Be films coated with 5 \AA\ of Au.    
 
 In addition to inducing spin-orbit scattering, an Au overlayer can affect the film in three other ways.  First,  gold atoms sitting on the Be surface may increase the interface scattering rate, and correspondingly the room temperature resistance of the film.  Second, because the overlayer marginally increases the overall conductive thickness of the Be film, it may reduce $e-e$ interaction effects.  This mechanism can, in fact, produce an {\it inverse} proximity effect in highly disordered superconducting films that are near the S-I transition \cite{Dynes2002}.  Finally, at sufficiently high Au coverages one would expect that any local superconducting amplitude will be suppressed by the overlayer via the {\it standard} proximity effect in the Cooper limit \cite{deGennes1964}.  

Be/Au bilayers of varying Au thickness were prepared by e-beam evaporation in an initial vacuum of $\sim0.1$ $\mu$Torr. All of the depositions were made on fire polished glass substrates held at 84 K.  First a Be film with thickness $\sim18$ \AA\ was deposited at a rate of 1.4 \AA/s, then a Au overlayer was deposited at 0.1 \AA/s without breaking the vacuum.  The morphology of the Be and Be/Au films was probed via atomic force microscopy and found to be very smooth and homogenous,  with no evidence of islanding or granularity \cite{Adams1998}.  Magnetotransport measurements up to 9 T were made in a dilution refrigerator with a base temperature 50 mK using a standard 4-wire dc I-V method with probe currents of a few nano-amps.  This has the advantage of avoiding the undesirable phase shifts that are often encountered when using ac techniques on high resistance samples.  The films in the dilution unit were aligned with the magnetic field via an {\it in situ} mechanical rotator. Hall effect measurements were made on samples that were vapor cooled down to 1.8 K in magnetic fields up to 9 T via a Quantum Design PPMS.  

\begin{figure}
\includegraphics[scale=0.6]{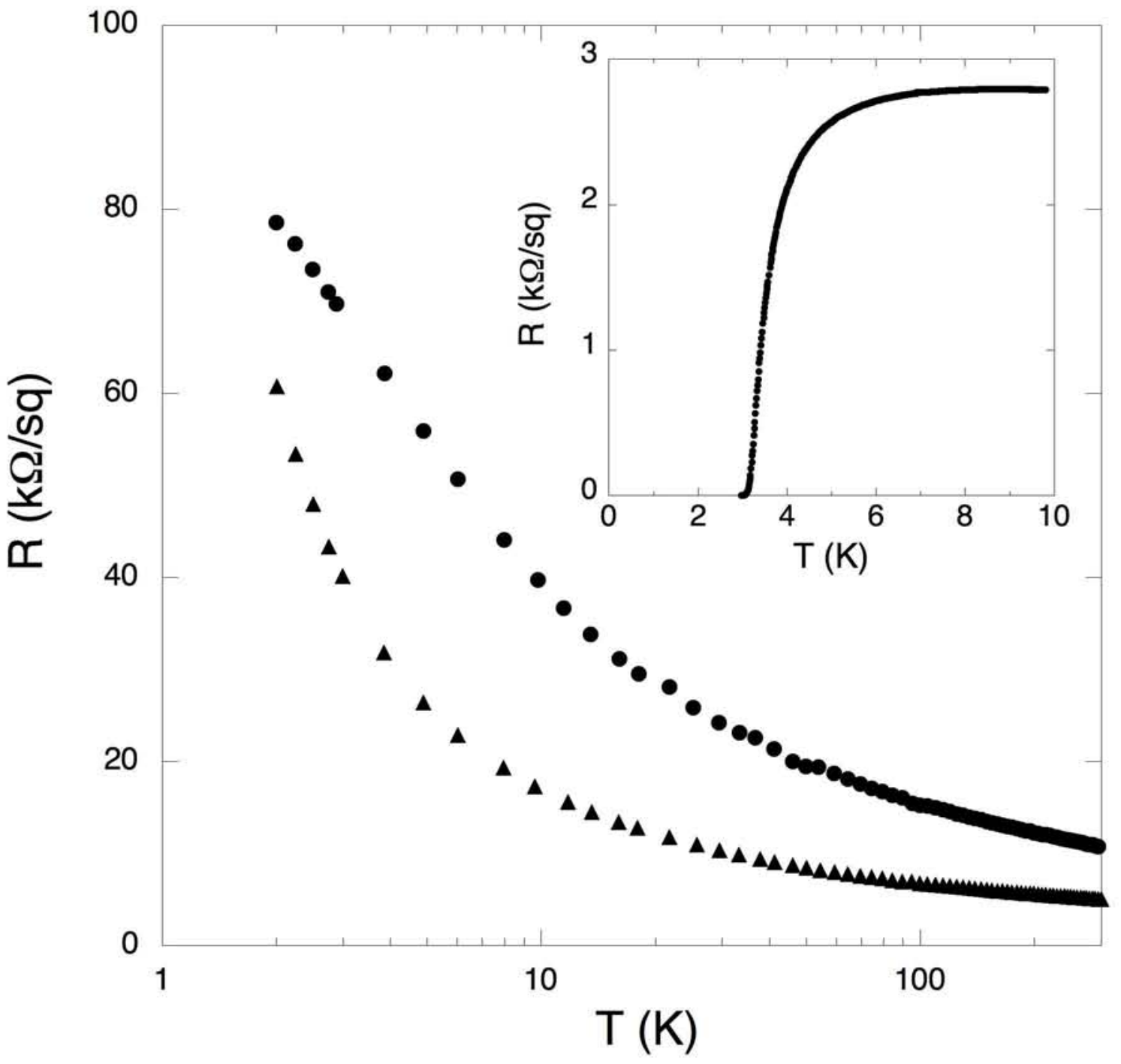}
\caption{\label{RT} Temperature dependence of the sheet resistance of a 18 \AA\  Be film (triangles) and a 18 \AA/0.6 \AA\  Be/Au bilayer (circles). Inset: superconducting transition in a 23 \AA\ Be film, with $T_c\sim3.5$ K.} 
\end{figure}

In the inset of Fig.\ \ref{RT}  we show the resistive superconducting transition from a $23$ \AA\ Be film with $R\ll R_Q$.  The midpoint transition temperature $T_c\sim3.5$ K is about a factor of five higher than what we typically observed in previous studies.  We believe that the increase in $T_c$ is due to the fact that the depositions were made at significantly lower chamber pressures, 0.1 $\mu$Torr versus 0.4 $\mu$Torr in Refs.\ \onlinecite{Adams1998,Wu2006}.  Films with Be thickness below $\sim17-18$ \AA\ did not superconduct and were found to be strongly insulating, as can be seen in the main panel of Fig.\ \ref{RT}, where we plot the temperature resistance of a 18 \AA\ Be film along with a 18 \AA/1.0 \AA\ Be/Au bilayer.  Note that the room temperature resistance of the bilayer is significantly higher than that of the uncoated Be film at room temperature.   However, as we will show below, its low-temperature correlation energy is much smaller than its uncoated counterpart.   

\begin{figure}
\includegraphics[scale=0.6]{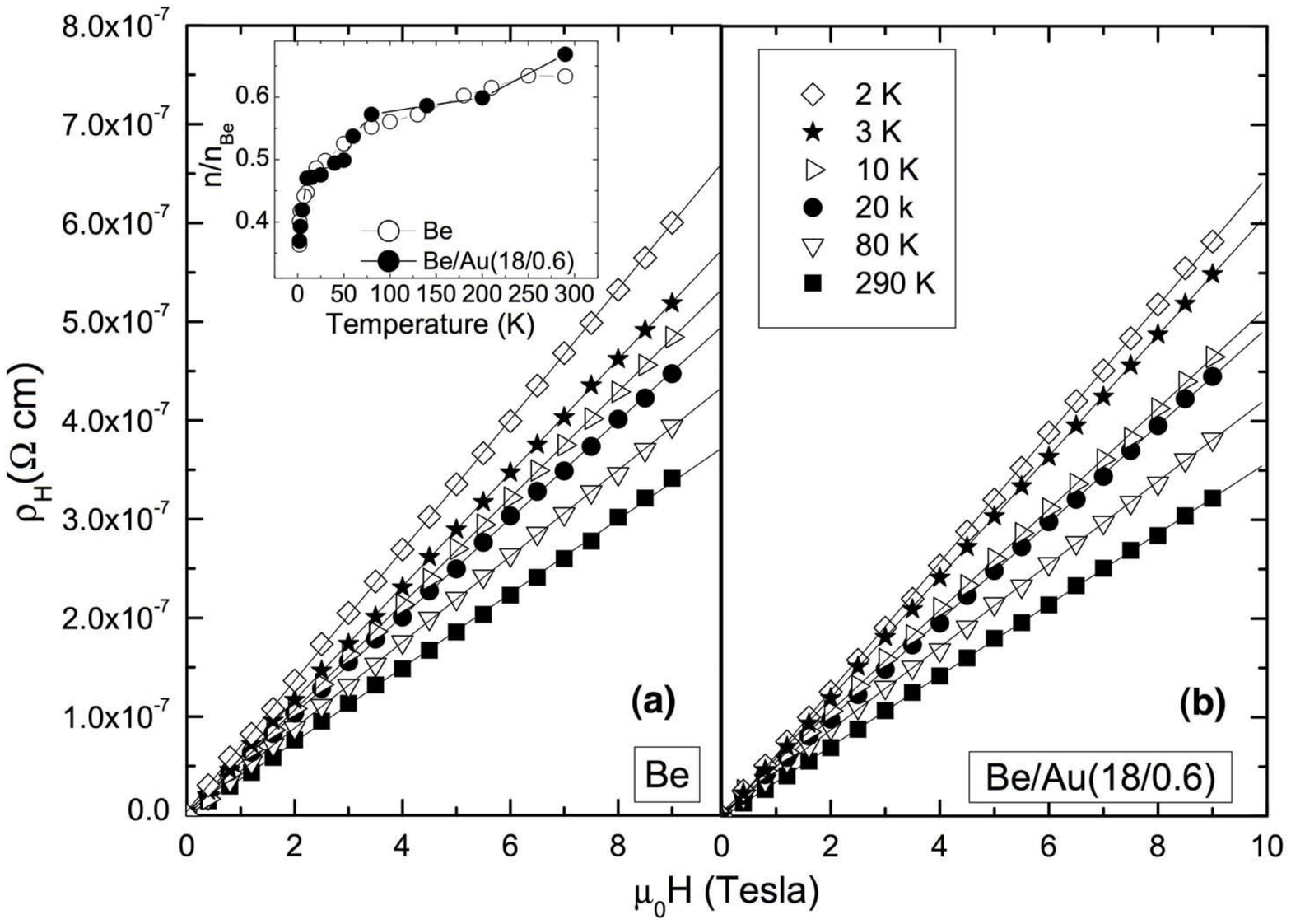}
\caption{\label{Hall} Field dependence of the Hall resistivity of a 18 \AA\ Be film and a 18 \AA/0.6 \AA\ Be/Au bilayer.  Inset: Hall carrier density normalized by the bulk Hall carrier density of polycrystalline Be, plotted as a function of temperature. }
\end{figure}

Beryllium has a low density of states near the Fermi energy, so it is useful to measure the film carrier density to insure that there is not a significant Au doping effect in the bilayers.  In Fig.\ \ref{Hall} we present Hall effect measurements on the two films with similar thicknesses to the ones used in Fig.\ \ref{RT}.  The Hall angle is small in metal films, so a large drive current must be used, $\sim10\ \mu$A, to obtain a reasonable signal.  As can be seen in the figure, the Hall voltage of both films was linear, temperature dependent, and hole-like.   Hall measurements in bulk, polycrystalline beryllium also show hole-like transport with a Hall carrier density $n_{Be}=2.6\ \rm{x}\ 10^{22}$ cm$^{-3}$.  The 2p orbitals in Be are believed to have a high mobility and thus dominate the Hall response.   In the inset of Fig.\ \ref{Hall} we plot the carrier density, as obtained from the Hall data, as a function of temperature, where we have normalized the density by its bulk value.  Note that at high temperatures the carrier density in both films is about a factor of two lower than the bulk value.  However, as temperature is lowered below 20 K, a precipitous drop in carrier density occurs. This drop roughly coincides with the rapid increase in resistance.  We believe that the shoulder in the carrier density curve represents a depletion of the density-of-states associated with the emergence of the correlation energy $T_o$.  The S-I transition, in fact, is largely governed by the competition between the correlation fluctuations represented by $T_o$ and the superconducting fluctuations represented by the low-disorder limit of $T_c$.  

\begin{figure}
\includegraphics[scale=0.6]{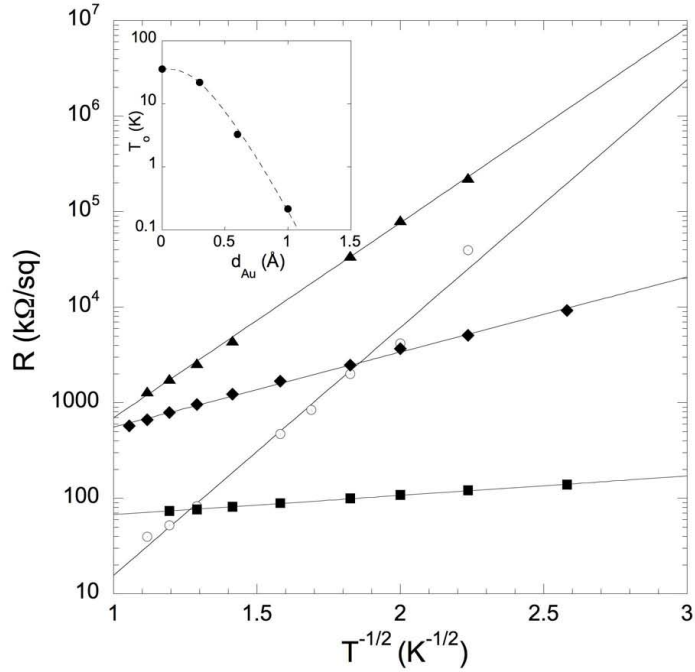}
\caption{\label{Scaling} Low temperature scaling behavior of the sheet resistance of Be/Au bilayers with varying Au coverage, $d_{Au}$.  The Be thickness of each film is 18\AA.  Circles: $d_{Au}=0.0$ \AA. Triangles: $d_{Au}=0.3$ \AA.  Diamonds: $d_{Au}=0.6$ \AA. Squares: $d_{Au}=1.0$ \AA.  The solid lines represent least-squares fits to \req{MVRH} from which the correlation energy $T_o$ was obtained.  Inset: correlation energy as function of Au coverage.  The dashed line is provided as a guide to the eye.}
\end{figure}

In Fig.\ \ref{Scaling} we show the low-temperature scaling behavior of four Be/Au bilayers, each having a Be thickness of 18 \AA\ but varying Au thickness.   Each of the samples exhibits transport that
is of the modified variable range hopping form
\begin{equation}
\label{MVRH}
R(T)=R_o\exp{(T_o/T)^\nu},
\end{equation}   
where $R_o$ is a constant, $T_o$ is the correlation energy, and the hopping exponent $\nu\sim0.5$, consistent with that reported by Butko {\it et al.} \cite{Butko2000b} on Be films with low-temperature resistances much larger than $R_Q$.    The solid lines represent a least-squares fit to \req{MVRH} where $T_o$ and $R_o$ were varied.   Clearly the correlation energy of the bilayers decreases with increasing Au coverage, suggesting that the Au is suppressing $e-e$ interactions.  In the inset we plot $T_o$ as a function of the Au thickness $d_{Au}$.  It is interesting to note that the magnitude of $T_o$ in the present Be films is more than an order of magnitude larger than what we previously reported in Be films.  However, the latter films displayed a much lower characteristic superconducting energy scale $T_c\sim0.7$ K on the superconducting side of the S-I transition, which suggests that $T_o$  and $T_c$ are correlated.

\begin{figure}
\includegraphics[scale=0.7]{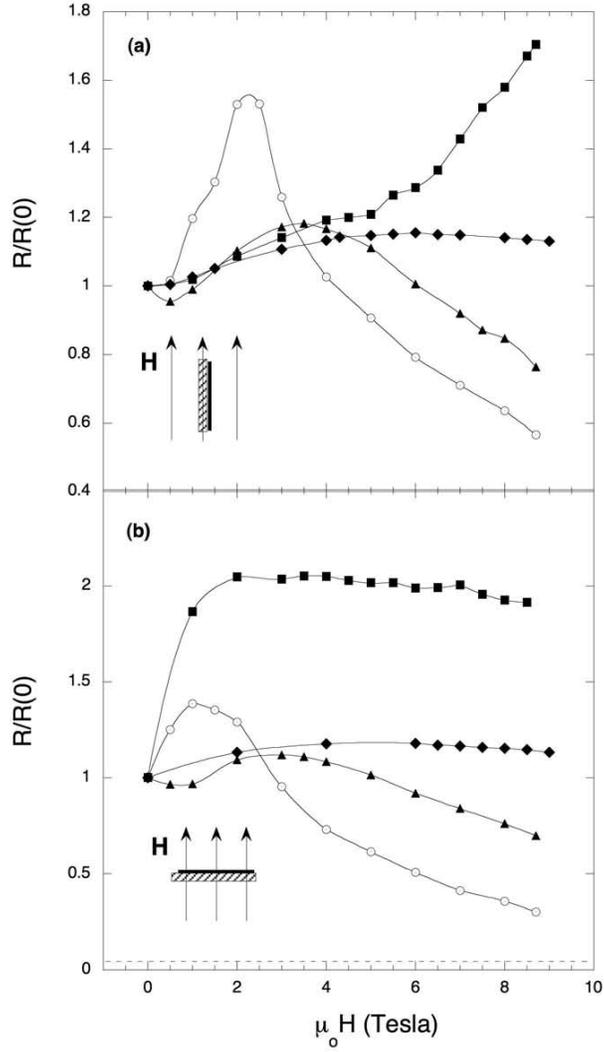}
\caption{\label{MRdata} The normalized resistance of the Be/Au films in Fig.~(\ref{Scaling}) as a function of parallel (upper panel) and perpendicular magnetic field (lower panel).  The $d_{Au}=0.0$ curves were taken at 500 mK.  The other curves were taken at 400 mK.}
\end{figure}

In Fig.\ \ref{MRdata} we compare the perpendicular and parallel-field MR of the Be/Au bilayers in Fig.\ \ref{Scaling}. The data, which have been normalized by the zero field resistance, were taken at $400 - 500$ mK in order to circumvent the long, non-exponential, relaxations that hinder measurements below 100 mK.   The open circle symbols correspond to the uncoated 18 \AA\ Be film which displays  the previously reported low-field positive MR followed by a multi-fold negative MR \cite{Butko2000,Butko2000b}.  The dashed line near the x-axis in panel (b) of Fig.\ \ref{MRdata} corresponds to $R_Q$ for the Be film.  The MR appears to be asymptotic to $R_Q$, in accord with the high-field quantum metal phase \cite{Butko2001}.  In contrast, the overall scale of the MR for the $d_{AU}=0.3$ \AA\ bilayer is somewhat diminished and almost completely quenched in the $d_{Au}=0.6$ \AA\ bilayer.   Both of these samples have low temperature sheet resistances $R\gg R_Q$ and correlation energies $T_o\gg T$, as is the case for the uncoated film.    Because of this we conclude that the MR is being modified by the spin-orbit scattering and not the lowering of $T_o$, for instance.  Note that the MR peaks move to substantially higher field with increasing Au coverage in the $d_{Au}=0.0,0.3, {\rm and}\ 0.6$ curves.  The $d_{Au}=1.0$ curves display the largest MR anisotropy, but this may be a consequence of the fact that $T_o\sim T$ for this sample.  Nevertheless, the high-field perpendicular MR is only weakly negative for the highest Au coverage bilayer, while the parallel MR maximum, if it exists, lies beyond 9 T.  

The films are sufficiently thin so as to rule out an orbital response in the parallel field orientation.  Note that the structure and magnitude of the $d_{Au}= 0, 0.3, {\rm and}\ 0.6$ MR curves in Fig.\ \ref{MRdata} are relatively insensitive to field orientation.  This suggests that the MR in the $T_o>>T$ limit is dominated by electron Zeeman splitting.  If, indeed, the zero-field insulating phase is mediated by localized Cooper pairs, {\it i.e.}\ a Bose insulator \cite{Fisher}, then the MR of the Be films can be attributed to pair-breaking via Zeeman splitting of the localized Cooper pairs.  To account for the field range of the MR, one must assume that there is a rather broad distribution of local pair binding energies.  Since the Zeeman-mediated critical field of low SOS superconductors is simply proportional to the superconducting gap \cite{Wu2006}, it is natural to assume that the local pair-breaking field in the insulating phase will also be proportional to the local energy gap.  However, in the presence of SOS the Zeeman critical field of a superconductor can be much larger than that of the zero-SOS case.  Therefore, it is not surprising that the field scale for local pair-breaking in the Be/Au bilayers is much higher than the that of pure Be films.

In summary, we show that, by coating high resistance Be films with sub-monolayer coverages of Au, a large interface SOS scattering rate can be induced.  Though the Au overlayer tends to lower the correlation energy $T_o$ of insulating Be films, it does not change the scaling exponent nor the overall character of the S-I transition.  Consequently the zero-field transport characteristics of the Be films with and without Au look very similar.  However, the spin response of the system is radically altered by the SOS.  We speculate that the quantum metal phase emerges once the zero-field Bose insulator phase is quenched via Zeeman splitting of localized Cooper pairs.     	     
 
We gratefully acknowledge enlightening discussions with James Valles and Ilya Vekhter. This work was supported by the DOE under Grant No.\ DE-FG02-07ER46420.  DPY acknowledges support from the NSF under Grant No.\ DMR-0449022.

\bibliographystyle{apsrev}
\bibliography{SIT_Localization}


\newpage

\end{document}